\documentclass[german,english]{bvm2020}

\def\articlenumber{0000}

\date{}
\title{Invertible Neural Networks for Uncertainty Quantification in Photoacoustic Imaging}

\titlerunning{Invertible Neural Networks in Photoacoustic Imaging}

\author{Jan-Hinrich~Nölke$^{1,2}$, Tim~Adler$^{1,3}$, Janek~Gröhl$^{1}$, Thomas Kirchner$^{4}$, Lynton~Ardizzone$^{5}$, Carsten~Rother$^{5}$, Ullrich~Köthe$^{5}$, Lena~Maier-Hein$^{1,3,6}$}

\authorrunning{Nölke et al.}

\institute{%
$^1$Division of Computer Assisted Medical Interventions, German~Cancer~Research~Center, Heidelberg, Germany\\
$^2$Faculty of Physics and Astronomy, Heidelberg University, Heidelberg, Germany\\
$^3$Faculty of Mathematics and Computer Science, Heidelberg~University,~Heidelberg,~Germany\\
$^4$Institute of Applied Physics, University of Bern, Bern, Switzerland\\
$^5$Visual Learning Lab, HCI, IWR, Heidelberg, Germany\\
$^6$Medical Faculty, Heidelberg University, Heidelberg, Germany\\}

\email{j.noelke@dkfz-heidelberg.de}

\begin{document}

%
\selectlanguage{english}

\maketitle

\begin{abstract}
Multispectral photoacoustic imaging (PAI) is an emerging imaging modality which enables the recovery of functional tissue parameters such as blood oxygenation. However, the underlying inverse problems are potentially ill-posed, meaning that radically different tissue properties may - in theory - yield comparable measurements. In this work, we present a new approach for handling this specific type of uncertainty by leveraging the concept of conditional invertible neural networks (cINNs). Specifically, we propose going beyond commonly used point estimates for tissue oxygenation and converting single-pixel initial pressure spectra to the full posterior probability density. This way, the inherent ambiguity of a problem can be encoded with multiple modes in the output. Based on the presented architecture, we demonstrate two use cases which leverage this information to not only detect and quantify but also to compensate for uncertainties: (1) photoacoustic device design and (2) optimization of photoacoustic image acquisition. Our \textit{in silico} studies demonstrate the potential of the proposed methodology to become an important building block for uncertainty-aware reconstruction of physiological parameters with PAI.

\end{abstract}
\section{Introduction} \label{Introduction}
Photoacoustic Imaging (PAI) is an emerging medical imaging modality, which enables the recovery of optical tissue properties with a "light-in-sound-out" approach \cite{zackrisson_light_2014}. The key idea is to use the initial pressure distribution $p_0$ to determine physiological tissue properties like blood oxygenation $sO_2$. However, the non-linear effect of the light fluence makes the optical inverse problem ill-posed \cite{yang_deep_2020}. This can potentially lead to ambiguous solutions of the tissue properties. Prior work has addressed related problems with different approaches to uncertainty quantification \cite{tarvainen_bayesian_2013,tick_image_2016,grohl_confidence_2018,godefroy_solving_2020}, yet, explicitly representing ambiguities by full posterior distributions has not been attempted in the context of machine learning-based image analysis.

In this work, we address this gap in the literature with conditional invertible neural networks (cINNs) \cite{ardizzone_guided_2019}. In contrast to conventional neural networks, the INN architecture enables the computation of the full posterior density function (rather than a simple point estimate), which naturally enables the encoding of various types of uncertainty, including multiple solutions (modes).
The contribution of this paper is two-fold: (1) We adapt the concept of cINNs to the specific problem of quantifying tissue parameters from PAI data. (2) We demonstrate the value of our approach with two use-cases, namely photoacoustic device design and optimization of photoacoustic image acquisition.

\section{Materials and methods}
\begin{figure}
    \centering
    \includegraphics[width=1.0\textwidth]{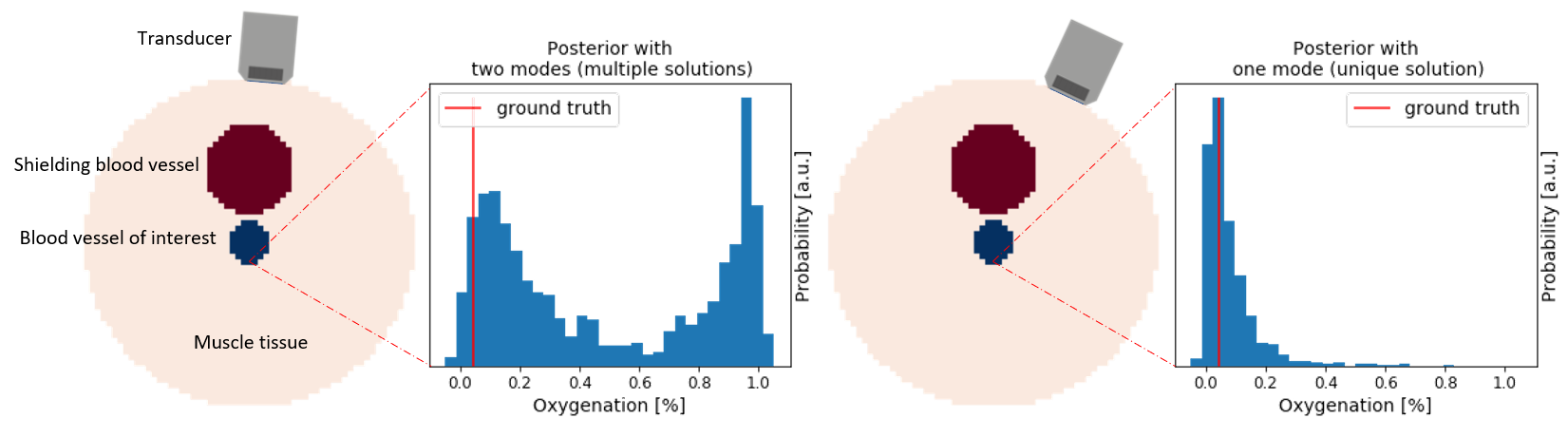}
    \caption{\textit{In silico} setting illustrating how slight changes in the PAI probe pose can resolve model ambiguity (training on S2b, see sec.~\ref{Experiments}). Left: the posterior corresponding to a pixel of interest features two modes. Right: Owing to an improved acquisition pose, the same pixel features a uni-modal posterior.}
    \label{fig:multimodes}
\end{figure}

\subsection{Virtual photoacoustic imaging environment}
The virtual environment that was created for testing the proposed approach to uncertainty quantification is based on a digital PAI device. With it 3D representations of the optical and acoustic properties of tissue can be generated, which are used to simulate synthetic PAI data for a given probe design, pose and ground truth tissue properties. The data is simulated using the Monte Carlo eXtreme (MCX) framework \cite{fang_monte_2009}. In the implementation for this study, each simulation is performed with $10^7 \ts {\rm photons}$ originating from a pencil-like source and a grid spacing of $0.34 \ts {\rm mm}$. Each volume is simulated at 26 equidistant wavelengths between $700 \ts {\rm nm}$ and $950 \ts {\rm nm}$. 

\subsection{Approach to uncertainty quantification} \label{INN}

Our architecture builds upon the cINN architecture proposed in \cite{ardizzone_guided_2019}.
cINNs transform an input $x$ (in our case blood oxygenation) given a conditioning input $y$ (in our case a single-pixel initial pressure spectrum) to a latent variable $z$. Maximum likelihood (ML) training ensures $z$ to be distributed according to a standard normal distribution. During inference time, because of the invertible architecture, we can sample the latent distribution and, given a conditioning input $y$, generate a conditional probability distribution $p(x|y)$. 
The specific architecture implemented in this work consists of 20 blocks, each of which comprises a random (but fixed) permutation and a conditional generative flow (GLOW) coupling block \cite{kingma_glow_2018} (two fully connected layers of size 512 and rectified linear unit (ReLU) activations). During training, we apply normally distributed random noise with $\sigma=0.001$ to the normalized input and $\sigma=0.1$ to the conditioning input. The models are trained for 60 epochs with the AdamW optimizer and weight decay of $0.01$. We start with a learning rate of $10^{-3}$ and reduce it by a factor of 10 after epoch 40 and 50.     
 
 In order to automate the detection of multi-modal posteriors we introduce a multi-mode score.  We perform kernel density estimation on the posterior samples with 21 different bandwidths between $0.01 \ts {\rm p.p.}$ and $0.1 \ts {\rm p.p.}$. The score is then simply the fraction of estimates with more than one maximum relative to all estimates.

\subsection{Experiments} \label{Experiments}
\begin{figure}
    \centering
    \includegraphics[width=1\textwidth]{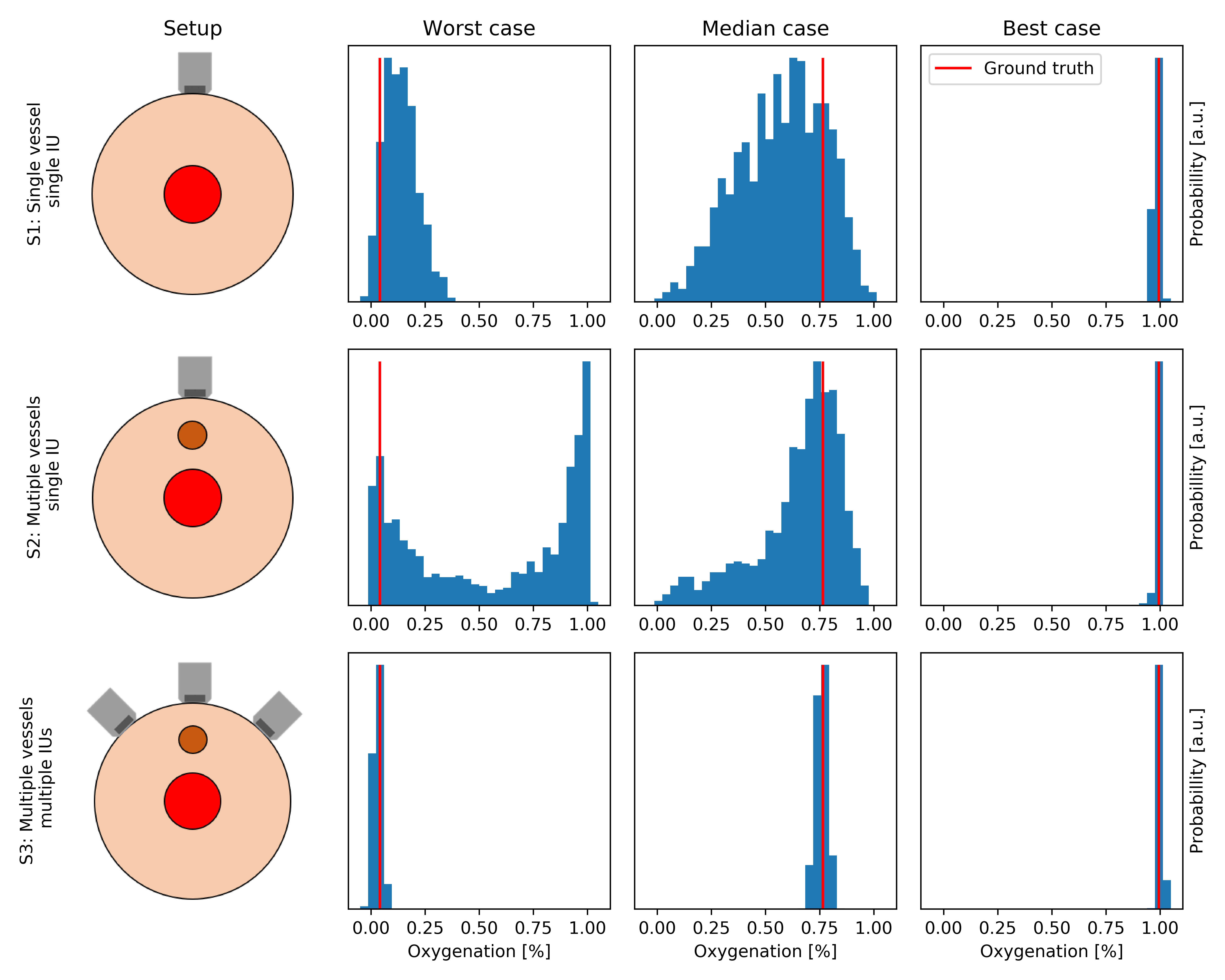}
    \caption{Worst, median and best case with respect to the IQR of S2 for three investigated settings. In contrast to a single vessel scenario (top), ambiguities are likely to occur in a multi-vessel scenario (middle) when using a single light source. These can be compensated for with multiple light sources (bottom).}
    \label{fig:histo}
\end{figure}

The purpose of our experiments was to (1) validate the proposed approach to uncertainty quantification in PAI and to (2) showcase use cases that leverage the posteriors to not only detect and quantify uncertainties but to compensate for them. To this end, we generated four different settings.
\begin{description}
\item[S1: Single vessel, single illumination unit (IU):] Images (probabilistically) generated for this setting comprise a tube of muscle tissue with $2 \ts {\rm cm}$ diameter as background with a blood oxygenation uniformly drawn between $0$ and $1$. In the center, a blood vessel with a radius uniformly drawn between $1 \ts {\rm mm}$  and $3 \ts {\rm mm}$ and oxygenation between 0 and 1 is placed. A single illumination source is used.

\item[S2: Multiple vessels, single IU:] The setting S1 is enhanced by introducing an additional blood vessel randomly placed between the light source and the central vessel of interest. This vessel also has a radius between $1 \ts {\rm mm}$ and $3 \ts {\rm mm}$ and an oxygenation between 0 and 1. Fig.~\ref{fig:multimodes} illustrates the basic setup of the phantoms.

\item[S2b: Multiple vessels, shifted single IU:] This setting is identical to S2, but the scene is illuminated from two additional angles ($\pm$45\textdegree). We used the three different illumination setups as independent samples leading to a three times bigger data set. This setting (S2b) was exclusively used to generate Fig.~\ref{fig:multimodes}, i.\,e.\ to demonstrate the effect of probe position on the resulting posterior.

\item[S3: Multiple vessels, multiple IUs:] The setting uses the same data as S2b, but we concatenate the three spectra from the different illumination setups (thus simulating a complex device with three illumination units/detectors) leading to a conditioning input dimension of $3\cdot26$. Fig.~\ref{fig:histo} gives an overview about the settings S1-S3.

\end{description}

We simulated 2,000 volumes for each of the settings and trained cINN models as described in sec.~\ref{INN} on each of them with 85\% of the data. The remaining 15\% of the data was used for testing.
To validate the accuracy of the posteriors, we computed the  calibration curves for scenarios S1-S3 as proposed in \cite{ardizzone_analyzing_2019}.
We further processed the results corresponding to all settings to analyze the capability of our method to reveal ambiguous problems (represented by multiple modes) and to determine the effect of device pose and design.

\section{Results}

\begin{figure}
	\centering
	\includegraphics[width=1\textwidth]{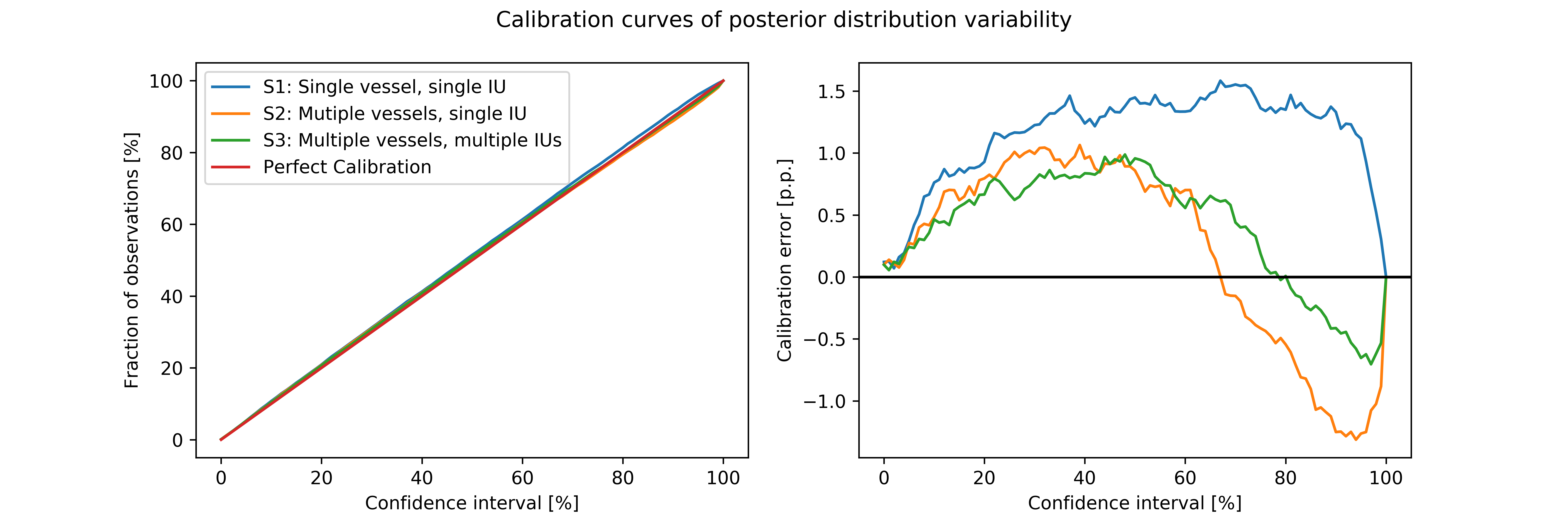}
	\caption{Calibration curves of the posterior distributions of the settings S1-S3 as described in sec.~\ref{Experiments}. Fraction of observations (left) and calibration error (right) as a function of confidence interval on the test set.}
	\label{fig:cal}
\end{figure}

As can be seen in Fig.~\ref{fig:cal} all calibration curves are very close to the identity (median calibration error < 1.5 p.p.). This implies that the width of our posteriors is reliable. For the setting with single vessel and single illumination (S1) the model is slightly underconfident.

\begin{figure}
	\centering
	\includegraphics[width=1\textwidth]{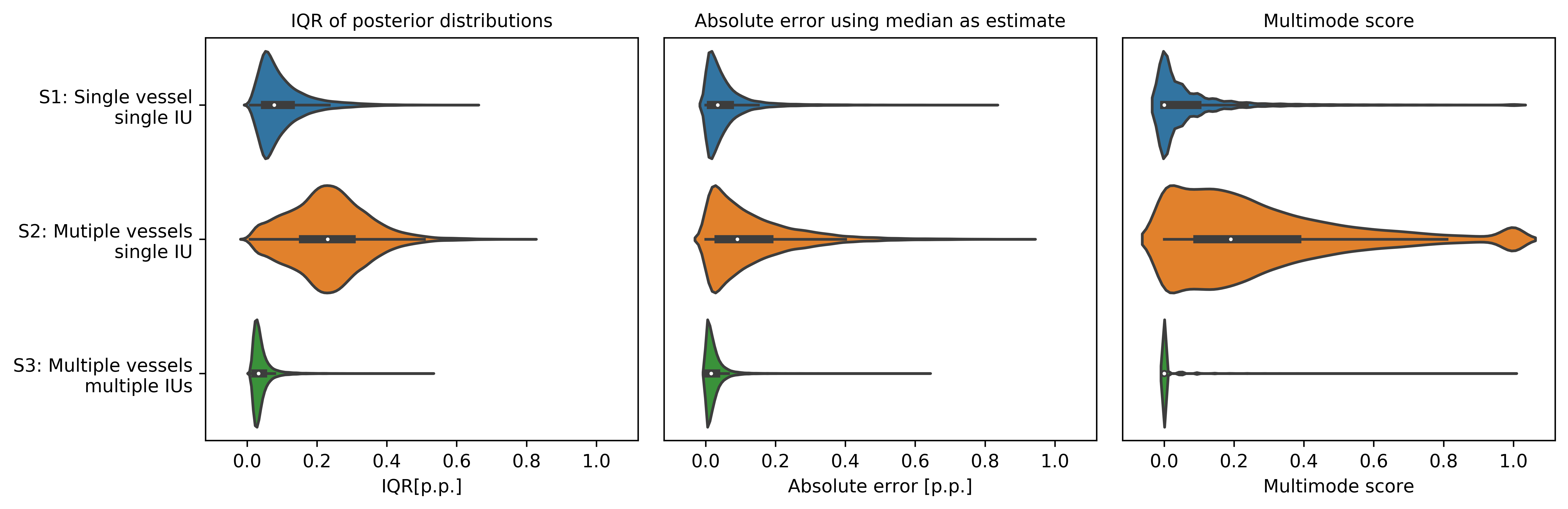}
	\caption{The violin plots show the interquartile range (IQR) of the posterior distribution on the test set, the absolute error when using the median as estimate and the multi-mode score, introduced in sec.~\ref{INN}. We differentiate between the settings S1-S3 as described in sec.~\ref{Experiments}.}
	\label{fig:violin}
\end{figure}

 In Fig.~\ref{fig:violin} we compare the distribution of IQRs, absolute errors and the multi-mode score for the scenarios S1-S3 described in sec.~\ref{Experiments}. Our results demonstrate that not only the accuracy but also the likelihood for ambiguity of the problem depends crucially on the characteristics of the probe (e.g. number of illumination/detection units). For all three metrics, the performance for the setting with multiple vessels, but only one illumination (S2) is clearly the worst. In particular, this setting includes a non-negligible fraction of multi-modal posteriors. 

Fig.~\ref{fig:multimodes} and Fig.~\ref{fig:histo} further show that the accuracy in a given pixel depends crucially on the pose and the illumination geometry of the PAI device. Moreover, ambiguity of the inverse problem can be potentially resolved by performing the acquisition from a different position/angle or by using a multiple illumination setting (S3). Our approach could thus serve as a basis for optimizing the measurement process and photoacoustic device design. 

\section{Discussion}
To our knowledge, this is the first work exploring the concept of INNs in the context of PAI. Specifically, we have demonstrated the capabilities of cINNs to represent and quantify uncertainties in the context of physiological parameter estimation. Based on our initial experiments, we believe that our approach could also serve as a basis for optimizing PAI probe design and image acquisition.

With regard to device design, this work is similar to that proposed by  Adler et al.~\cite{adler_uncertainty-aware_2019} in the context of multispectral optical imaging. However, this work differs in that we used cINNs instead of the original INN architecture, which comes along with several major advantages, including (1) no zero-padding needed, leading to smaller network size, (2) maximum likelihood training and (3) no hyperparameters in the loss function.

Our findings with respect to device design are in line with Shao et al.~\cite{shao_estimating_2011} where a multi-illumination setup is suggested to improve image reconstruction. Our initial experiments indicate that our method may help in the optimization of the acquisition process. As a next step, our approach has to be extended such that it not only shows the current ambiguities, but also proposes possible poses to resolve them. This might be achieved through the application of reinforcement learning. 

In conclusion, we have demonstrated the potential of cINNs to reconstruct tissue parameters from PAI data while systematically representing and quantifying uncertainties. Future work will focus on translating the work to a real setting. 

\ack{This project has received funding from the European Union’s Horizon 2020 research and innovation programme through the ERC starting grant COMBIOSCOPY under grant agreement No. ERC2015-StG-37960. }

\bibliographystyle{bvm2020}

\bibliography{2650}
\marginpar{\color{white}E\articlenumber} 
\end{document}